%% file: main.tex
\documentclass[preprint,journal]{vgtc}

\onlineid{1563}

\vgtccategory{Research}

\newcommand{\oursystem}{SVS}
\newcommand{\linear}{\textsc{Linear}}
\newcommand{\summaryguided}{\textsc{Summary-guided}}
\newcommand{\sourceguided}{\textsc{Source-guided}}

\title{Spatial Visual Analytics for Multi-Document Summary Verification}

\author{
Jiahao Xu, Wei Liu, Yang Liu, Eric Krokos, Kirsten Whitley, Xuan Wang, Rebecca Faust, and Chris North
}

\authorfooter{
  \item Jiahao Xu, Wei Liu, Yang Liu, Xuan Wang, and Chris North are with Virginia Tech.
  E-mail: \{jiahao, wliu3, yangliu07, xuanw, north\}@vt.edu.
  \item Eric Krokos and Kirsten Whitley are with the Department of Defense.
  \item Rebecca Faust is with Tulane University.
  E-mail: rfaust1@tulane.edu.
}

\abstract{Large language models increasingly generate summaries from collections of documents to support sensemaking and reporting, but verifying whether summary statements are grounded in source materials remains difficult. In multi-document summarization (MDS), evidence is distributed across many source documents and may be incomplete, conflicting, or missing. 
We present Summary Verification Space (\oursystem{}), a visual analytics system for verifying multi-document summaries through spatial document organization and coordinated provenance visualization. To support scalable verification, we investigate two alternative 2D canvas layouts: a \summaryguided{} layout that organizes documents by alignment with summary sentences, and a \sourceguided{} layout that arranges documents by semantic similarity. Coordinated provenance visualization then makes relationships among summary content, source documents, and supporting evidence explicit, enabling users to trace support, contradiction, and missing evidence during verification. A task-driven usage scenario illustrates an auditing workflow in which users use the layouts to locate relevant documents and coverage gaps, then inspect linked claims and source evidence to make their own grounding judgments. In a comparative study with provenance held constant, both spatial layouts improved aggregate accuracy and reduced workload relative to a linear baseline, with the clearest gains on relevance tasks. The \summaryguided{} layout provides the strongest overall balance of accuracy, efficiency, confidence, and workload.
}


\teaser{
  \centering
  \includegraphics[width=0.82\linewidth]{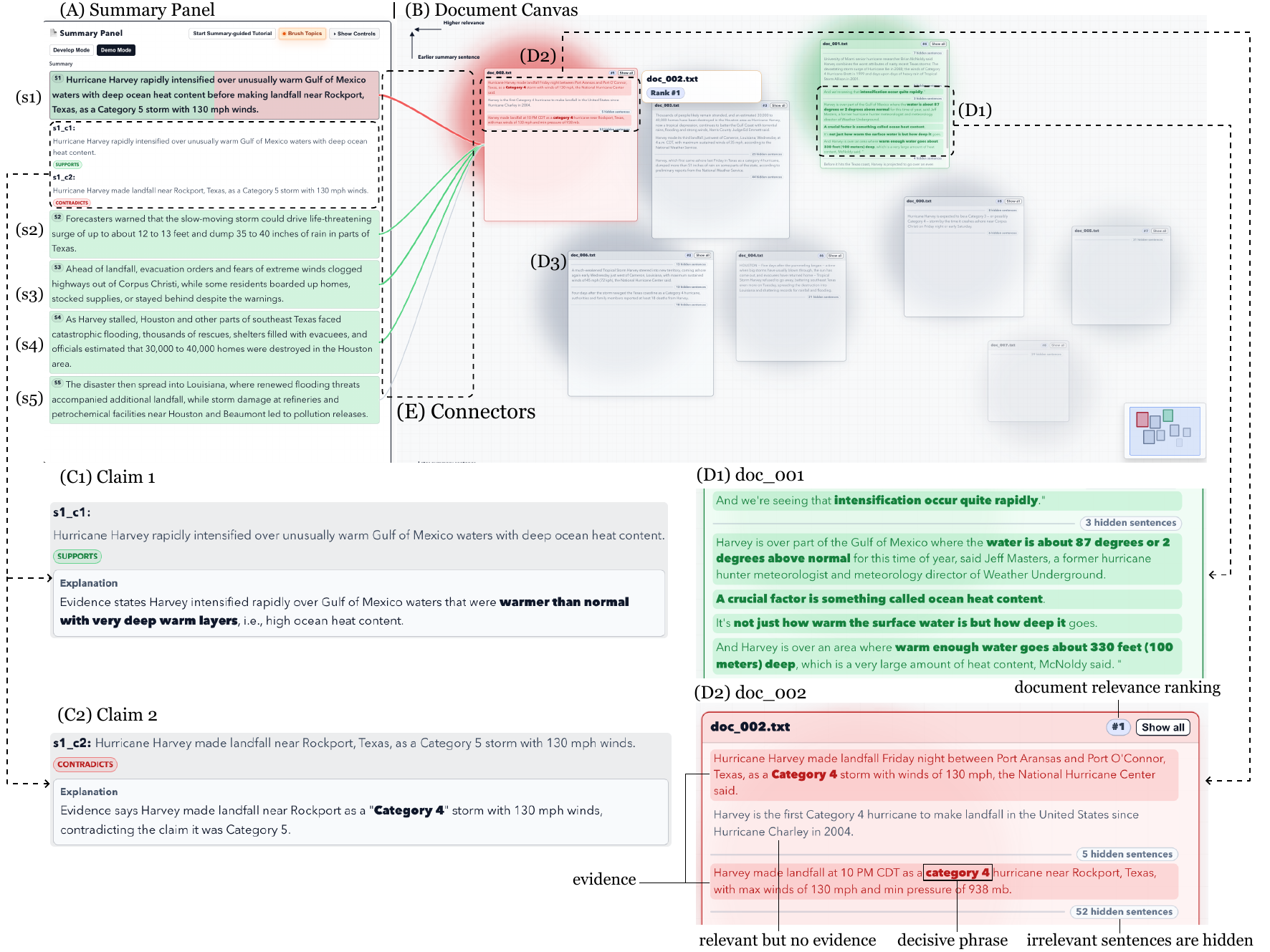}
  \caption{
  	Summary Verification Space (\oursystem{}). The summary panel (A) presents sentences (S1--S5), decomposed claims (C1--C2), inference judgments, and explanations. The document canvas (B) arranges source cards (D1--D3), highlights supporting or contradicting evidence, compresses irrelevant content, and links summary content to evidence with connectors (E).
  }
  \label{fig:teaser}
}

\graphicspath{{figures/}{vendor/ieee-tvcg-2026/}{./}}

\usepackage{tabu}
\usepackage{booktabs}
\usepackage{lipsum}
\usepackage{mwe}
\usepackage{ccicons}
\usepackage{cuted}
\usepackage{placeins}

\usepackage{mathptmx}
\usepackage{amssymb}
\usepackage{amsmath}

\usepackage[most]{tcolorbox}
\newtcolorbox{takeawaybox}{
    colback=blue!5,
    colframe=blue!40,
    boxrule=0.5pt,
    arc=1pt,
    left=4pt,
    right=4pt,
    top=3pt,
    bottom=3pt,
    boxsep=0pt,
    before skip=0.35em,
    after skip=0.35em
}
\newenvironment{compactquote}{\begin{list}{}{\leftmargin=1.2em\rightmargin=1.2em\topsep=0.15em\partopsep=0pt\parsep=0.15em\itemsep=0pt}\item\relax\itshape}{\end{list}}

\definecolor{supportLowColor}{RGB}{190, 80, 70}
\definecolor{supportMediumColor}{RGB}{210, 145, 45}
\definecolor{supportHighColor}{RGB}{55, 126, 84}
\definecolor{supportEmptyColor}{RGB}{255, 255, 255}
\newcommand{\harveyball}[2]{\tikz[baseline=-0.65ex]{\fill[supportEmptyColor] (0,0) circle (0.75ex);\ifnum#1=2 \fill[#2] (0,0) -- (90:0.75ex) arc[start angle=90,end angle=270,radius=0.75ex] -- cycle;\else\ifnum#1=3 \fill[#2] (0,0) circle (0.75ex);\fi\fi\draw[black,line width=0.35pt] (0,0) circle (0.75ex);}}
\newcommand{\supportlow}{\harveyball{0}{supportLowColor}}
\newcommand{\supportmedium}{\harveyball{2}{supportMediumColor}}
\newcommand{\supporthigh}{\harveyball{3}{supportHighColor}}

\begin{document}

\firstsection{Introduction}\label{sec:intro}
\maketitle

Large language models (LLMs) are increasingly used to generate summaries to support sensemaking and decision-making workflows~\cite{zhang2025systematic, dhaini2024explainability}. By condensing large information spaces into concise narratives, summaries help users quickly grasp key topics and relationships without reading extensive source materials~\cite{allahyari2017brief}. However, automatically generated summaries may contain unsupported statements, overlook important information, or present incomplete or conflicting interpretations~\cite{huang2021factual}. As a result, users must determine whether summary content is factually grounded in the underlying sources.

This need makes summary verification a critical step in summary-supported analysis~\cite{gabriel2021go, maynez2020faithfulness, kryscinski2020evaluating}. For example, during a literature review, a researcher using an automatically generated survey summary must verify whether key claims are supported by findings reported in individual research papers, identify studies that present conflicting results, and check whether important prior work or alternative viewpoints have been overlooked. Visual analytics offers a promising approach for verifying summaries by enabling users to externalize relationships between summaries and sources~\cite{gotz2009characterizing, wang2019vizseq, chan2020melody, endert2012semantic}. 
However, existing approaches are less effective in multi-document summary (MDS) scenarios, where verification requires users to identify relevant evidence across many documents. 
Designing effective provenance visualization for verifying summaries against source documents in MDS is fundamentally challenged by two key verification characteristics: relevance and consistency~\cite{fabbri2021summeval}.

First, (\textbf{C1}) \textbf{relevance} poses a fundamental challenge in MDS verification. For any given summary, only a subset of the source collection is actually pertinent, yet that evidence may be scattered across many documents. Users must identify how relevance is distributed across the collection because the pattern of such distribution reveals if the summary covers important content from the sources. As document sets grow, this process becomes increasingly demanding, requiring repeated navigation and shifts of attention across sources. Supporting relevance is therefore a matter of helping users understand where pertinent evidence resides and how it is distributed at scale.~\cite{endert2012semantic, pirolli2005sensemaking}.

Second, (\textbf{C2}) \textbf{consistency} poses a complementary challenge. Even after relevant evidence has been identified, users must still determine whether that evidence collectively supports, contradicts, or fails to substantiate the summary. In MDS, such judgments rarely depend on a single passage; instead, they often emerge from synthesizing information across multiple claims, sentences, and documents, including cases where evidence is incomplete, complementary, or in conflict. Supporting consistency verification therefore requires more than tracing isolated evidence fragments. 

It requires helping users connect these evidence fragments to higher-level summary meaning~\cite{gotz2009characterizing, ragan2015characterizing, wang2019vizseq}.

To address these challenges, we present \oursystem{} (\cref{fig:teaser}), a visual analytics system for MDS verification that combines spatial document organization for \textbf{(C1) relevance} with integrated provenance visualization for \textbf{(C2) consistency}. 
For \textbf{C1}, \oursystem{} centers on an interactive 2D canvas workspace that spatially organizes source documents. We identify two alternative layouts from different perspectives. The \summaryguided{} layout arranges documents by their relationships to summary sentences, making summary--document relevance immediately visible and helping users identify which parts of the collection are most pertinent to a verification target. The \sourceguided{} layout instead places documents in a semantic embedding space, and positions summary sentences as visual anchors, so that users can interpret relevant documents within the broader structure of the source collection. We conduct a user study to compare these layouts, identify design tradeoffs, and measure improvement with respect to a baseline 1D layout used in existing systems.

\oursystem{} supports \textbf{C2} through an integrated provenance representation that links summary content to its evidence across the collection. Summary sentences are first decomposed into verifiable claims, and each claim is connected to localized evidence passages retrieved from the source documents. These provenance links are overlaid directly on the workspace, connecting summary and source documents within a coordinated view. This design allows users to move seamlessly between global relevance patterns and local evidence inspection, supporting judgment of whether summary content is supported, contradicted, or insufficiently grounded.
Our paper offers three contributions:
\begin{itemize}[noitemsep, topsep=0pt, leftmargin=*]
    \item Two alternative spatial document layout strategies designed to support multi-document summary verification.
    \item An integrated provenance visualization design that supports reasoning about summary–source relationships across granularities.

    \item A usage scenario and user study demonstrating the value of spatial organization and provenance visualization for MDS verification.
\end{itemize}

\section{Related Work}
\label{sec:relatedwork}
\subsection{Visualizing Summary Verification}

Early verification approaches focused on surface coverage assessment, commonly operationalized through lexical overlap metrics such as ROUGE~\cite{lin2004rouge}. Visualization research at this stage primarily supported similarity inspection workflows, employing ranked document views, token-level highlighting, and similarity heatmaps to help users examine correspondence between summaries and source texts at the lexical level~\cite{stasko2008jigsaw, hearst1995tilebars}.

As summarization verification research evolved toward semantic grounding, newer evaluation methods introduced question-answering reconstruction~\cite{durmus2020feqa, wang2020asking}, semantic similarity scoring~\cite{gao2020supert, zhang2019bertscore}, and natural language inference (NLI) models~\cite{laban2022summac, zha2023alignscore} to assess whether summary claims are supported by source evidence. Visualization work in this phase increasingly emphasized exploration of semantic relationships, moving beyond lexical inspection to support reasoning about evidence relevance at scale~\cite{liu2018nlize, strobelt2018s}. For example, SummVis~\cite{vig2021summvis} provides coordinated lexical and semantic visualizations that align model outputs, reference summaries, and source documents, enabling detailed inspection of factual consistency and abstraction patterns in generated summaries. Similarly, Summary Explorer~\cite{syed2021summary} supports comparative analysis of summarization systems through interactive relevance ranking and similarity-based grouping views that reveal relationships among summaries and source evidence. Other interpretable summarization interfaces such as RTSUM~\cite{cho2024rtsum} visualize salience signals at multiple granularities, including sentence- and phrase-level importance, helping users understand how salient textual units contribute to summary generation. These techniques improve faithfulness by enabling users to trace how individual summary statements relate to supporting passages and navigate semantic relationships across heterogeneous document collections. More recently, the emergence of LLM-as-a-Judge paradigms~\cite{liu2023g, li2024llms} has further expanded opportunities for visualization. By producing structured verification outputs such as entailment labels, confidence scores, and natural language rationales, LLM-based evaluation methods allow interfaces to incorporate claim-level annotations, explanation panels, and evidence highlighting views that support user interpretation of automated judgments~\cite{pan2025vis}.

Although these modern techniques improve faithfulness analysis for summary verification, they provide limited support for MDS verification, where evidence is distributed across many documents rather than localized within a single source or a small comparison set. In such settings, users must not only find relevant evidence, but also understand how that evidence is organized across the collection and whether multiple sources collectively support, contradict, or fail to ground the summary~\cite{deyoung2024multi, sarikaya2018design, ragan2015characterizing, vig2021summvis}.

These limitations motivate the development of visual analytics systems that integrate automated verification signals with coordinated provenance visualization and spatial reasoning support to enable comprehensive multi-document summary verification workflows.

\subsection{Spatial Document Organization}
Spatial organization is a well-established strategy for supporting sensemaking in complex analytic tasks. Prior research shows that analysts arrange documents and visual artifacts within spatial workspaces to externalize relationships, maintain contextual awareness, and reduce cognitive effort during reasoning \cite{andrews2010space, andrews2013impact}. Studies of large display analytics demonstrate that flexible spatial layouts can improve recall and facilitate reasoning about relationships in large data collections~\cite{isenberg2013visualization}.

Early work on spatial hypertext systems explored how informal spatial arrangements enable users to construct knowledge representations without rigid hierarchical structures~\cite{shipman1999formality}. These systems allowed users to group related documents, form visual clusters, and incrementally organize information in ways that reflect evolving analytic goals~\cite{marshall1995spatial}. Subsequent visual analytics tools extended these ideas by integrating spatial layouts with coordinated views~\cite{stasko2008jigsaw}, filtering mechanisms~\cite{cao2010facetatlas}, and interactive exploration techniques~\cite{collins2009docuburst}, enabling users to examine thematic structures, relevance relationships, and distributional patterns across document corpora. 

Recent research increasingly adopts spatial workspace paradigms to support exploration of semantic similarity and topic relationships across large textual corpora \cite{liu2024visualizing}. These systems employ clustering layouts, embedding projection techniques, and network representations to organize documents and model outputs within flexible visual canvases that facilitate exploratory reasoning~\cite{tang2024steering, tenney2020language}. For example, systems such as Sensecape~\cite{suh2023sensecape} support spatial grouping and iterative sensemaking across heterogeneous information sources, while ChainForge~\cite{arawjo2024chainforge} introduces node-based spatial workflows that enable analysts to compare and organize large language model outputs during prompt experimentation. Building on similar spatial reasoning principles, LADiCA~\cite{zhang2025ladica} provides a 2D canvas for arranging textual artifacts according to semantic relationships, allowing users to construct spatial representations of evolving analytic insights. By enabling users to externalize semantic structures and iteratively refine spatial layouts, these approaches highlight the continued importance of spatial organization as a cognitive scaffold for reasoning over complex, large-scale text collections.

Despite these strengths, less is known about how spatial layouts can be combined with provenance visualization for systematic evidence assessment. This gap is especially important for multi-document summary verification, where users must synthesize distributed and potentially conflicting evidence across sources. We address this gap by combining spatial document layouts with coordinated provenance visualization, helping users navigate complex evidence structures and assess the faithfulness of generated summaries.

\section{Design Goals and System Overview}

Our analysis of MDS verification suggests two design goals. 

\textbf{DG1: Scalable Overview of Relevant Evidence.}
In MDS verification, only a subset of documents and passages is pertinent to any given summary sentence or claim, yet that evidence may be distributed across many long sources. Users need support for understanding how relevant evidence is distributed across the collection because the pattern of such distribution reveals if the summary covers important content from the source.
A verification interface should therefore provide an overview through spatial organization that reduces search and navigation burden while preserving meaningful relationships among documents.
This design goal primarily addresses the challenge of \textbf{(C1) relevance}.

\textbf{DG2: Explicit Representation of Consistency Relationships.}
In MDS verification, users must do more than identify relevant passages; they must determine whether distributed evidence collectively supports, contradicts, or fails to substantiate a summary statement. 
These judgments are often made between source documents and individual summary sentences, or even atomic claims within a sentence, rather than solely at the level of the summary as a whole~\cite{laban2022summac, zhang2024finegrainednaturallanguageinference}.
The interface should therefore integrate verification outputs into a coherent provenance representation that links localized evidence passages to atomic claims, summary sentences, and source documents. This design goal primarily addresses the challenge of \textbf{(C2) consistency}.

Following these goals, we designed \oursystem{}, a visual analytics system for supporting the verification of MDS through integrated provenance visualization and spatial document organization. The system interface is organized around two coordinated components: a summary panel (\cref{fig:teaser}A) that presents the generated summary together with sentence- and claim-level verification results, and a 2D canvas (\cref{fig:teaser}B) that displays source documents as interactive document cards. 
More implementation details are provided in the \href{https://osf.io/6j5c4/overview?view_only=5903d7db4ce24045974f7d8ae911fd26}{Supplemental Appendix~C}. \Cref{sec:spatial,sec:provenance,sec:usage,sec:userstudy} describe the spatial layouts, provenance visualization, usage scenarios, and comparative study.

\begin{figure*}
    \centering
    \includegraphics[width=\linewidth]{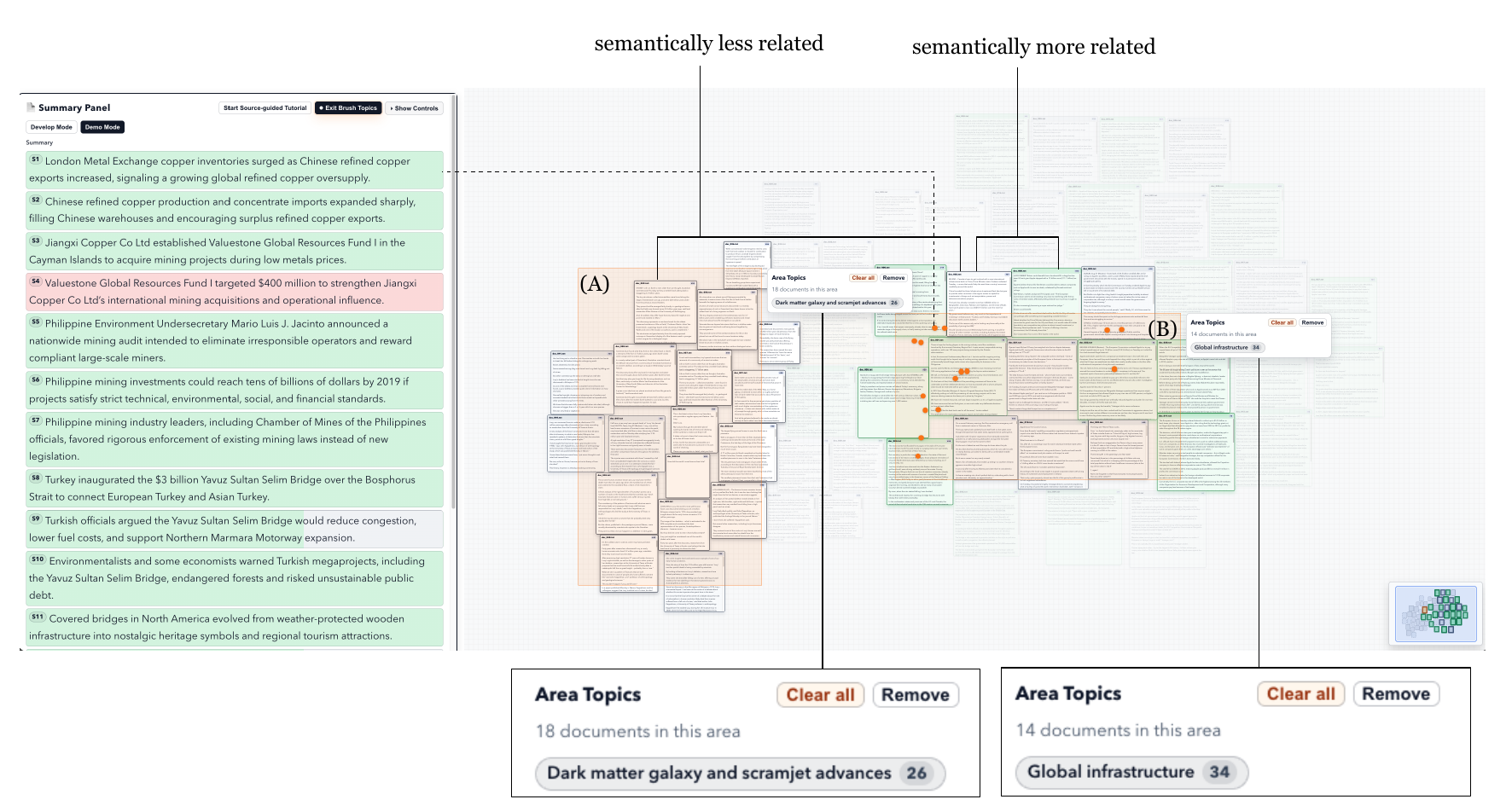}

    \caption{\sourceguided{} layout. 
    Documents form semantic clusters, while summary anchors show where evidence is concentrated. Brushed regions expose local topics, revealing that region A is largely uncovered by the summary.}
    \label{fig:sourcelayout}
    \vspace{-1em}
\end{figure*}

\section{Spatial Organization of Source Documents}
\label{sec:spatial}

To support \textbf{DG1}, \oursystem{} organizes source documents spatially to provide a scalable overview of relevant evidence. 
We investigate two alternative grounding strategies for spatially organizing source documents: a novel \summaryguided{} projection (\cref{fig:teaser}), which derives document positions from their alignment with summary structure, and a \sourceguided{} semantic projection (\cref{fig:sourcelayout}), which follows conventional embedding-based dimensionality reduction to preserve semantic relationships among source documents.

\subsection{\summaryguided{} Layout}
The \summaryguided{} layout organizes source documents according to how they contribute to the structure of the summary. When assessing whether summary statements are well supported, users need to identify which documents are important overall and determine which documents contribute evidence to specific parts of the summary. The layout therefore arranges documents using two spatial ordering signals.

\subsubsection{Horizontal Ordering: Overall Document Relevance}
The horizontal axis encodes overall document relevance to the summary. Intuitively, this dimension answers the question: Which documents matter more for verifying this summary? During verification, users often prioritize documents that contain strong supporting or contradicting evidence. To reflect this need, relevance is determined by the strongest sentence-level relationship observed between a document and any summary sentence.

For each document $d$, the system constructs a similarity matrix $M_d \in \mathbb{R}^{n_d \times K}$, where $n_d$ is the number of sentences in the document and $K$ is the number of summary sentences. Each entry 
\begin{equation}
\label{eq:sentence-summary-similarity}
M_d[i,j] = sim(s_i^d,t_j)
\end{equation}
represents the semantic similarity between document sentences $s_i^d$ and summary sentence $t_j$. The overall relevance score for document $d$ is then defined as 
\begin{equation}
R_d = \max_{i,j}|M_d[i,j]|.
\end{equation}

Documents are sorted in descending order of $R_d$ and mapped to evenly spaced horizontal positions on the canvas. This places documents containing decisive evidence, either strong support or strong contradiction, closer to the summary panel. For instance, in \cref{fig:teaser}B, D1 is positioned to the left of D2 and D3, indicating that it is more important to the summary overall. Horizontal position therefore acts as a prioritization cue for which documents users should inspect first.

\subsubsection{Vertical Ordering: Alignment with Summary Structure}
The vertical axis encodes where a document’s strongest positive evidence occurs within the summary sequence. This dimension answers the question: Which part of the summary does this document primarily relate to? It complements horizontal ordering because two documents can be equally relevant overall but correspond to different summary sentences or claims. Without this cue, users would still need to search within a ranked set to determine where each document fits in the verification workflow.

To estimate this alignment, the system computes for each summary sentence $j$ the strongest positive similarity observed in the document:
\begin{equation}
w_d(j) = \max_i \max(M_d[i,j], 0).
\end{equation}

These values form a non-negative distribution describing how evidence from document $d$ is distributed across summary sentences. The document's alignment center is then defined as the weighted mean summary index:
\begin{equation}
\mu_d = \frac{\sum_{j=1}^K j\cdot w_d(j)}{\sum_{j=1}^K w_d(j)}.
\end{equation}

Documents are sorted by $\mu_d$ and assigned evenly spaced vertical positions. Documents whose strongest evidence relates to earlier summary content appear higher in the layout, while those aligned with later summary regions appear lower. For example, in \cref{fig:teaser}B, D1 and D2 align more strongly with earlier summary content, such as S1, while D3 aligns more strongly with later summary content, such as S3. By following the top-to-bottom order of the summary, the vertical arrangement turns the layout into a sentence-indexed evidence map, helping users separate evidence for different summary parts and notice where some summary regions have little document support.

\subsection{\sourceguided{} Layout}

The \sourceguided{} layout organizes source documents according to their semantic relationships within the collection rather than their relevance to summary order. During verification, users often need to understand how documents cluster around shared topics, where evidence is concentrated across the source set, and how individual summary sentences relate to these semantic neighborhoods. The layout therefore proceeds in two stages: it first projects documents into a space, and then places summary sentences within this document space using semantic signals.

\begin{figure*}
    \centering
    \includegraphics[width=\linewidth]{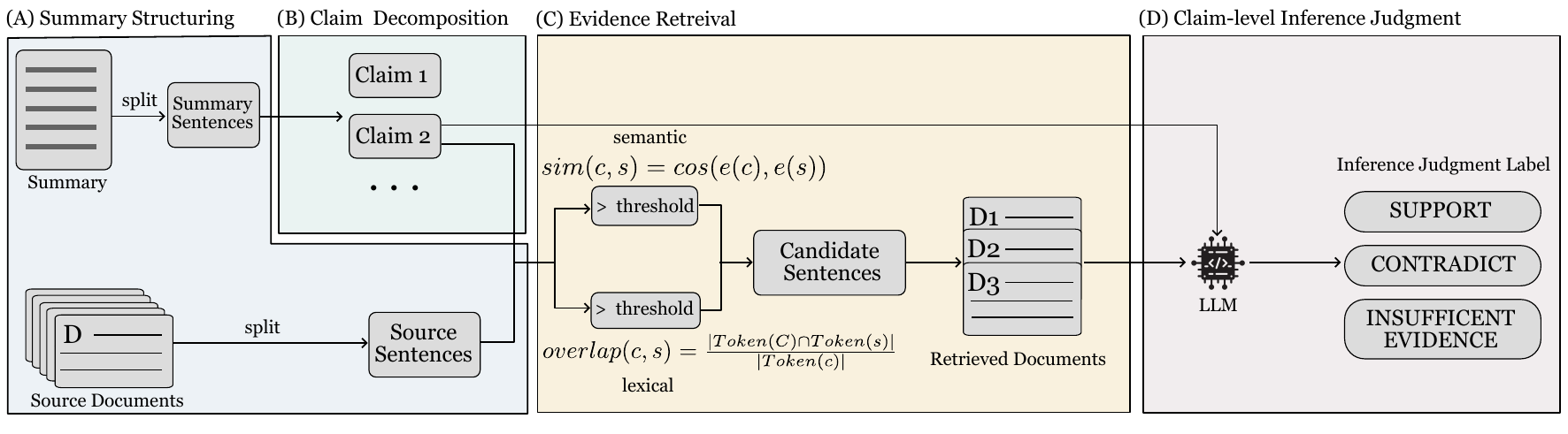}
    \caption{Verification pipeline for provenance analysis. SVS transforms summary content into structured provenance through four stages: summary structuring, claim decomposition, evidence retrieval, and claim-level inference judgment. Candidate evidence is identified using a hybrid semantic and lexical filter, and an LLM-based judge assigns \textsc{Supports}, \textsc{Contradicts}, or \textsc{Insufficient Evidence} labels.}
    \label{fig:pipeline}
    \vspace{-1em}
\end{figure*}

\subsubsection{Document Organization: Semantic Projection of Sources}

The \sourceguided{} layout begins by organizing documents according to overall semantic similarity. It represents each document using a whole-document embedding and projects the document collection into two dimensions via UMAP~\cite{mcinnes2018umap} (or any other commonly-used projection algorithm like t-SNE~\cite{van2008visualizing}). Documents that are similar in overall content are therefore placed near one another, while documents that are less relevant are positioned farther apart. This allows document clusters to reveal coherent topical neighborhoods within the source set.

Unlike the \summaryguided{} layout, the spatial axes in this mode do not carry explicit meaning. Instead, the layout is interpreted relationally: document position is meaningful primarily with respect to nearby documents. The distance between two projected documents,
\begin{equation}
\|\mathbf{z}_i-\mathbf{z}_j\|,
\end{equation}
approximates their semantic similarity. As a result, the projection forms a semantic map of the source collection, where users read proximity as document similarity and clusters as source-level topic structure.

\subsubsection{Summary Sentence Placement: Evidence Anchoring}

After document positions are established, summary sentences are placed into the projection using the same sentence-level similarity matrices $M_d$ from \cref{eq:sentence-summary-similarity}. For each summary sentence $t_j$, the system uses the document-level contribution signal $w_d(j)$ to represent how strongly document $d$ contributes evidence to that sentence. These contribution strengths are then normalized into weights, and each summary sentence is placed at the weighted average of document positions:
\begin{equation}
\mathbf{p}_j=\sum_d \alpha_d(j)\mathbf{z}_d,
\end{equation}
where $\alpha_d(j)$ denotes the normalized contribution weight of document $d$ for summary sentence $t_j$. This process places each summary sentence near the documents that contribute most strongly to it. The resulting sentence anchors indicate where evidence for each summary statement is concentrated in the semantic document map. Anchors located between clusters suggest that evidence is distributed across multiple document groups rather than concentrated in a single source region. For example, as shown in \cref{fig:sourcelayout}, summary sentence anchors are concentrated in region B, corresponding to documents related to global infrastructure, while region A, associated with dark matter and scramjet advances, remains largely separated from the anchors. This spatial pattern suggests that evidence for the summary is concentrated within one topical region of the collection, whereas another coherent topic cluster receives little or no coverage.

\section{Coordinated Provenance Visualization}
\label{sec:provenance}

To support MDS verification, \oursystem{} represents provenance as coordinated relationships among summary and source documents at multiple levels. These relationships are derived through a pipeline that decomposes summary content, retrieves evidence, and assigns claim-level inference judgments. The coordinated provenance is then visualized from the pipeline outputs. We first describe the verification pipeline, then the visual encodings and coordination mechanisms.

\subsection{Verification Pipeline for Provenance Analysis}

As shown in \cref{fig:pipeline}, \oursystem{} derives the provenance shown in the interface through four stages: summary structuring, claim decomposition, evidence retrieval, and claim-level inference judgment.

\textbf{Summary structuring.}
Given a summary, the backend first segments it into an ordered list of summary sentences using spaCy~\cite{honnibal2020spacy} sentence segmentation. Source documents are segmented into sentences in the same way, and all summary sentences and source sentences are embedded with the same embedding model. Using these embeddings, the system computes the sentence-level similarity matrices \(M_d\) defined in \cref{eq:sentence-summary-similarity}, where each entry measures the semantic similarity between a source sentence and a summary sentence. 

\textbf{Claim decomposition.}
Each summary sentence is then decomposed into a set of atomic claims using an LLM. An atomic claim is a single verifiable proposition extracted from the sentence. The decomposition process enforces that claims remain self-contained and preserves attribution and hedging, so the resulting claims can be judged independently without changing the meaning of the original sentence. 

\textbf{Evidence retrieval.}
For each atomic claim, the system retrieves candidate evidence from source-document sentences using a hybrid lexical-semantic filter. Let \(c\) denote an atomic claim, \(s\) denote a source-document sentence, and \(e(\cdot)\) denote the sentence embedding. Each source sentence receives a semantic similarity score:
\[
\mathrm{sim}(c, s)=\cos(e(c), e(s)),
\]
and a lexical overlap score:
\[
\mathrm{overlap}(c,s)=\frac{|\mathrm{Token}(c)\cap \mathrm{Token}(s)|}{|\mathrm{Token}(c)|}.
\]
A candidate sentence is retained if either its semantic similarity exceeds a semantic threshold or its lexical overlap exceeds a lexical threshold. The retained candidates are then grouped by document, producing the retrieved documents.

\textbf{Claim-level inference judgment.}
Finally, each atomic claim is passed to an LLM-based judge together with the retrieved evidence
packets. For each claim, the model outputs one of three labels~\cite{laban2022summac}:
\begin{itemize}[leftmargin=*,nosep]
  \item \textcolor[RGB]{21, 128, 61}{\textsc{Supports}}: the retrieved evidence jointly establishes
the claim.
  \item \textcolor[RGB]{185, 28, 28}{\textsc{Contradicts}}: the retrieved evidence shows that at
least one material part of the claim is false.
  \item \textcolor[RGB]{71, 85, 105}{\textsc{Insufficient Evidence}}: the retrieved evidence is insufficient to support or contradict the full claim.
\end{itemize}
The model also returns reason phrases, retrieved documents, and cited support or contradiction spans.

Overall, this pipeline converts summary content into structured provenance that can be visualized throughout the interface, providing computational basis for the provenance encodings described next.

\subsection{Provenance Visualization}

\oursystem{} represents provenance across two coordinated views. The summary panel organizes provenance around summary sentences and their atomic claims, providing the primary entry point for verification. The document canvas grounds claim-level inference judgments in the source collection by showing how relevant evidence is distributed across documents and how each document relates to the current verification target.

\subsubsection{Summary Panel Provenance}

The summary panel serves as the primary entry point for provenance inspection. Provenance focus narrows progressively as users move from no active selection (\cref{fig:teaser}S2--S5), to a selected summary sentence (\cref{fig:teaser}S1), and then to an atomic claim (\cref{fig:teaser}C). This progression shifts the panel from summary-level overview to sentence-level inspection and finally to claim-level evidence assessment.

Each summary sentence is shown as its own sentence card. At the overview level, the card background encodes the sentence's provenance state. When the sentence contains claims with different inference judgments, the background becomes a left-to-right mixed gradient whose proportions reflect the distribution of \textcolor[RGB]{21, 128, 61}{\textsc{Supports}}, \textcolor[RGB]{185, 28, 28}{\textsc{Contradicts}}, and \textcolor[RGB]{71, 85, 105}{\textsc{Insufficient Evidence}} claims within that sentence. This makes sentence-level verification patterns visible without expanding the sentence. For example, in \cref{fig:teaser}, the background of the second sentence (S2) contains only green, indicating full support from source documents.

When a sentence is activated, the panel expands to reveal its atomic claims (\cref{fig:teaser}S1). Each claim is shown as a separate row with a status badge indicating one of three claim-level inference judgments: \textcolor[RGB]{21, 128, 61}{\textsc{Supports}}, \textcolor[RGB]{185, 28, 28}{\textsc{Contradicts}}, and \textcolor[RGB]{71, 85, 105}{\textsc{Insufficient Evidence}}. This decomposition exposes provenance at the level where individual verification decisions are made, rather than forcing judgment at the sentence as a whole.

Selecting a claim opens a compact explanation panel that presents the textual rationale for the assigned inference judgment (\cref{fig:teaser}C). Within this explanation, key phrases are bolded to isolate the most decisive parts of the justification. The summary panel therefore provides both a compact provenance overview and a claim-level explanation layer, while also establishing the provenance focus that drives the document-side encodings.

\subsubsection{Document Canvas Provenance}

Where the summary panel organizes provenance around verification targets, the document canvas grounds those targets in the source collection. Each source document is represented as a document card positioned within the spatial layout. Provenance at the document level is encoded through card color, which summarizes how the document relates to the current provenance focus. In addition, rank badges attached to card headers expose relevance ordering.

Within each card, provenance is further localized to source sentences (\cref{fig:teaser}D2). Retrieved evidence sentences are colored by evidence role so that supporting and contradicting evidence is visible directly in document context. Decisive phrases inside evidence sentences are bolded. Sentences that are relevant but not direct evidence remain visible with softer styling, and non-evidence regions are compressed into compact placeholders.

The canvas also provides provenance cues that operate across the collection. Connector lines (\cref{fig:teaser}E) link documents to summary sentences, with color indicating whether the document supports, contradicts, or insufficiently grounds each summary sentence. 

To support localized thematic inspection, users can brush regions of the canvas to open topic summaries for the selected neighborhoods (\cref{fig:sourcelayout}A and B). These summaries are derived by matching brushed documents to a precomputed topic corpus, allowing local topic composition to be inspected quickly and compared across regions.

Taken together, these encodings turn provenance into a spatial representation of evidence across documents. The canvas therefore supports not only local evidence reading within a document card, but also cross-document inspection of how support, contradiction, and insufficiency are distributed across the source collection.

\section{Usage Scenario}
\label{sec:usage}

\subsection{Relevance Task: Assess Summary Coverage}
\label{sec:relevance_usage}
In this section, we illustrate how \oursystem{} supports the relevance task of assessing summary coverage over a source collection~\cite{fabbri2021summeval}. We use a 60-document cluster from the WCEP dataset~\cite{gholipour-ghalandari-etal-2020-large}, containing 30 Business articles and 30 Science articles. The goal is to determine whether the summary covers the major topics represented in the source collection, a coverage question that is difficult to verify through prompting alone.

For example, a user could ask an LLM whether the summary covers the major topics in the source collection, but the resulting textual answer would still need to be audited against the collection. The key evidence is not only in individual cited passages, but also in how documents are distributed across topics, whether omitted content forms a substantial cluster rather than isolated examples, and whether the summary is concentrated around only part of the collection.

Because this evidence is distributed across 60 documents, manually comparing the summary against every source document is impractical. Instead, the \sourceguided{} layout provides an overview of how the summary is positioned within the semantic structure of the collection. As shown in \cref{fig:sourcelayout}, the summary sentence anchors are concentrated in one region of the canvas rather than distributed across the full set of documents. This pattern suggests that the summary is aligned with only part of the source collection.

To examine which topics are covered and which may be omitted, users can further inspect two regions of the canvas (\cref{fig:sourcelayout}A and B). Topic analysis shows that documents in Area B focus on "global infrastructure", whereas documents in Area A focus on "Dark matter galaxy and scramjet advances". The summary anchors lie primarily within Area B, while Area A remains disconnected to the summary. \oursystem{} further indicates that 26 documents belong to the Area A topic group, making it a substantial portion of the collection. Taken together, these cues suggest that the summary emphasizes the business topic while largely omitting a major science-related topic cluster. The \sourceguided{} layout therefore helps users identify important parts of the collection that are not adequately covered by the summary.

\subsection{Consistency Task: Verify Summary Sentence}
\label{sec:consistency_usage}
In this section, we illustrate how \oursystem{} supports the consistency task of verifying a summary sentence through coordinated provenance visualization. The scenario uses an eight-document hurricane cluster from the Multi-News dataset~\cite{fabbri-etal-2019-multi}, where the documents describe the same event but provide overlapping and sometimes conflicting evidence across sources. This walkthrough shows how provenance visualization in \oursystem{} helps trace support and contradiction, while the value of spatial document organization is evaluated later in \cref{sec:userstudy} through a comparative user study. The goal is to verify whether the following summary sentence is supported by the source documents, a consistency question that is difficult to audit through prompting alone:

\textit{\textbf{(S1)} ``Hurricane Harvey rapidly intensified over unusually warm Gulf of Mexico waters with deep ocean heat content before making landfall near Rockport, Texas, as a Category 5 storm with 130 mph winds.''}

For example, a user could ask an LLM whether S1 is supported by the sources, but a textual answer would still need to be audited against the evidence. S1 contains multiple factual claims, and the same source collection may support one part while contradicting another. The user would still need to check which claim each source addresses, whether the cited evidence justifies the assigned inference judgment, and where conflicting evidence appears across documents.

To make this mixed evidence structure inspectable, the verification begins from the summary panel. The background gradient of S1 immediately indicates that the sentence is only partially consistent with the source collection: one part is supported, while another part is contradicted. This sentence-level provenance cue tells the user that S1 should be inspected more closely rather than accepted as fully grounded.

Selecting S1 activates its associated provenance across the interface. Two source documents become emphasized on the canvas: \texttt{doc\_001} and \texttt{doc\_002}. Their contrasting encodings reveal different verification roles. \texttt{doc\_001} is surrounded by a green heatfield, indicating supporting evidence, while \texttt{doc\_002} is surrounded by a red heatfield, indicating contradicting evidence. Expanding S1 further reveals that it contains two atomic claims. At this stage, the provenance visualization already shows that the evidence for S1 is distributed across multiple documents rather than concentrated in one source.

To understand why S1 is partially verified, the claims are inspected individually. Selecting the first claim reveals its explanation in the summary panel and highlights the corresponding supporting evidence in \texttt{doc\_001}. Zooming into the document exposes the relevant sentences. These localized evidence cues confirm that Harvey intensified rapidly over Gulf of Mexico waters that were warmer than normal and had deep warm-water layers. In this way, the provenance visualization connects the \textsc{Supports} inference judgment for the claim to the specific source phrases.

The second claim can then be inspected in the same manner. Its explanation indicates that the inconsistency concerns Harvey's landfall strength. The summary states that Harvey made landfall as a \textit{Category 5} storm, whereas the source evidence identifies it as \textit{Category 4}. Inspecting \texttt{doc\_002} reveals two highlighted sentences that explicitly describe Harvey as a Category 4 hurricane at landfall near Rockport. These passages provide direct contradictory evidence and explain why S1 is not fully supported by the source collection.

The final judgment for S1 is therefore mixed. The claim about rapid intensification over unusually warm Gulf waters with deep ocean heat content is supported, while the claim about landfall as a Category 5 storm with 130 mph winds is contradicted by the source evidence. This scenario shows how provenance visualization in \oursystem{} supports consistency analysis by linking sentence-level verification status, claim-level inference judgments, document-level polarity, and localized evidence within a single workflow.

\section{Comparative User Study}
\label{sec:userstudy}
We conducted a comparative user study with 18 participants to evaluate how different document organization strategies affect user performance in MDS verification. Because provenance visualization was held constant across conditions, the study isolates the effect of spatial document organization. We investigate the following questions:
\begin{itemize}[noitemsep, topsep=0pt, leftmargin=*]
    \item \textbf{Q1}: Does spatial document organization improve user performance over a \linear{} baseline for multi-document summary verification?
    \item \textbf{Q2}: How do \summaryguided{} and \sourceguided{} layouts differ in supporting different verification tasks?
    \item \textbf{Q3}: How do users perceive and use the three organization conditions during verification?
\end{itemize}

\subsection{Study Design}
\subsubsection{Conditions}
All three conditions used the same summary panel and provenance visualization. They differed only in how source documents were organized within the workspace:
\begin{enumerate}[noitemsep, topsep=0pt, leftmargin=*]
    \item \linear{}: Documents are arranged in a fixed sequential order, without grouping based on summary structure or source relationships. 
    \begin{itemize}[noitemsep, topsep=0pt, leftmargin=*, label=--]
        \item It represents a common provenance pattern in mainstream commercial LLM assistants (e.g., NotebookLM~\cite{google2026notebooklm}) for summary verification: the summary is presented alongside a linear list of source materials that users can inspect.
    \end{itemize}
    \item \summaryguided{}: Documents are organized with respect to the summary, such that their positions reflect how they relate to different parts of the summary content.
    \item \sourceguided{}: Documents are organized according to inter-document relationships, such that documents with semantically similar or related source content are placed near one another.
\end{enumerate}

\subsubsection{Tasks}
Participants completed ten tasks spanning two categories: \textbf{relevance} (Tasks 1--7) and \textbf{consistency} (Tasks 8--10). Relevance tasks focused on identifying important topics and relevant documents, while consistency tasks focused on determining whether summary content was supported, contradicted, or unsupported by the source documents.

\paragraph{Relevance Tasks}
\begin{enumerate}[noitemsep, topsep=0pt, leftmargin=*]
    \item \textbf{Topic identification}: Identify which topics are discussed in the source documents.
    \item \textbf{Topic coverage estimation}: Estimate how many source documents discuss a given topic.
    \item \textbf{Summary emphasis comparison}: Determine whether the summary discusses one topic more than others.
    \item \textbf{Most relevant document to the summary}: Identify the source document that is most relevant to the summary as a whole.
    \item \textbf{Least relevant document to the summary}: Identify the source document that is least relevant to the summary as a whole.
    \item \textbf{Most relevant documents to a summary sentence}: For a given summary sentence, identify the top two documents that are most relevant to that sentence.
    \item \textbf{Most relevant summary sentence for a document}: For a given source document, identify the most relevant summary sentence.
\end{enumerate}

\paragraph{Consistency Tasks}
\begin{enumerate}[start=8,noitemsep, topsep=0pt, leftmargin=*]
    \item \textbf{Supporting documents for a summary sentence}: For a summary sentence, select all documents that contribute to its inference judgment label.
    \item \textbf{Contradicted claim identification}: Identify which claim is contradicted by the source documents.
    \item \textbf{Unsupported claim identification}: Identify which claim has neither supporting nor contradicting evidence (insufficient evidence) across the source documents.
\end{enumerate}

\subsubsection{Datasets}
\label{sec:datasets}
We used three 60-document collections derived from the WCEP dataset introduced in \cref{sec:usage}. Each collection combined documents from two topical categories to create mixed-topic verification settings:

\begin{enumerate}[noitemsep, topsep=0pt, leftmargin=*]
    \item \textbf{Arts and Environment}: 30 Arts and Culture documents and 30 Environment documents.
    \item \textbf{Disasters and Health}: 30 Disasters and Accidents documents and 30 Health and Medicine documents.
    \item \textbf{Crime and Sports}: 30 Crime and Law documents and 30 Sports documents.
\end{enumerate}

\subsubsection{Participants}
We recruited 18 participants from a university mailing list. The study protocol was approved by the Virginia Tech Institutional Review Board (Protocol No. 25-1176), and all participants provided informed consent before participating. All participants were able to read and understand English. During the study, participants completed questionnaires on a laptop while interacting with \oursystem{} on a 27-inch monitor.

\subsubsection{Procedure}
We used a within-subject design in which each participant completed all three interface conditions: \linear{}, \summaryguided{}, and \sourceguided{}. To mitigate condition-order effects, we counterbalanced the presentation order using all six permutations of the three conditions.

The dataset order was fixed as Dataset 1, Dataset 2, and Dataset 3 for all participants. As described in \cref{sec:datasets}, the three collections were carefully constructed to provide comparable mixed-topic verification settings, so dataset progression did not require additional counterbalancing. Each participant completed three rounds of tasks, one under each condition and on one dataset per round, with the mapping between condition and dataset determined by the assigned condition order.

Before starting each condition, participants were given a tutorial introducing the corresponding interface organization and interaction workflow. The tutorial was presented as an in-place guided overlay on top of the live workspace, allowing participants to learn the interface in the same environment used for the study tasks. After the tutorial, participants completed the assigned tasks for that condition. After all three conditions were completed, participants provided feedback on their experience using the system.

\subsubsection{Measures}
We collected both task-level and condition-level measures. At the task level, we recorded the time required to complete each task and participants' self-reported confidence in their answer on a 7-point Likert scale: 1 indicated \emph{very unconfident} and 7 indicated \emph{very confident}. 

After completing each condition, participants reported their perceived \textbf{mental demand}, reflecting the required cognitive effort, and \textbf{physical demand}, reflecting the required interaction effort. In addition, they provided free-text feedback describing which features of the condition they found helpful for completing the tasks. These responses were used to capture subjective workload and condition-specific impressions that were not reflected in task completion data alone.

\begin{figure}[t]
    \centering
    \includegraphics[width=\linewidth]{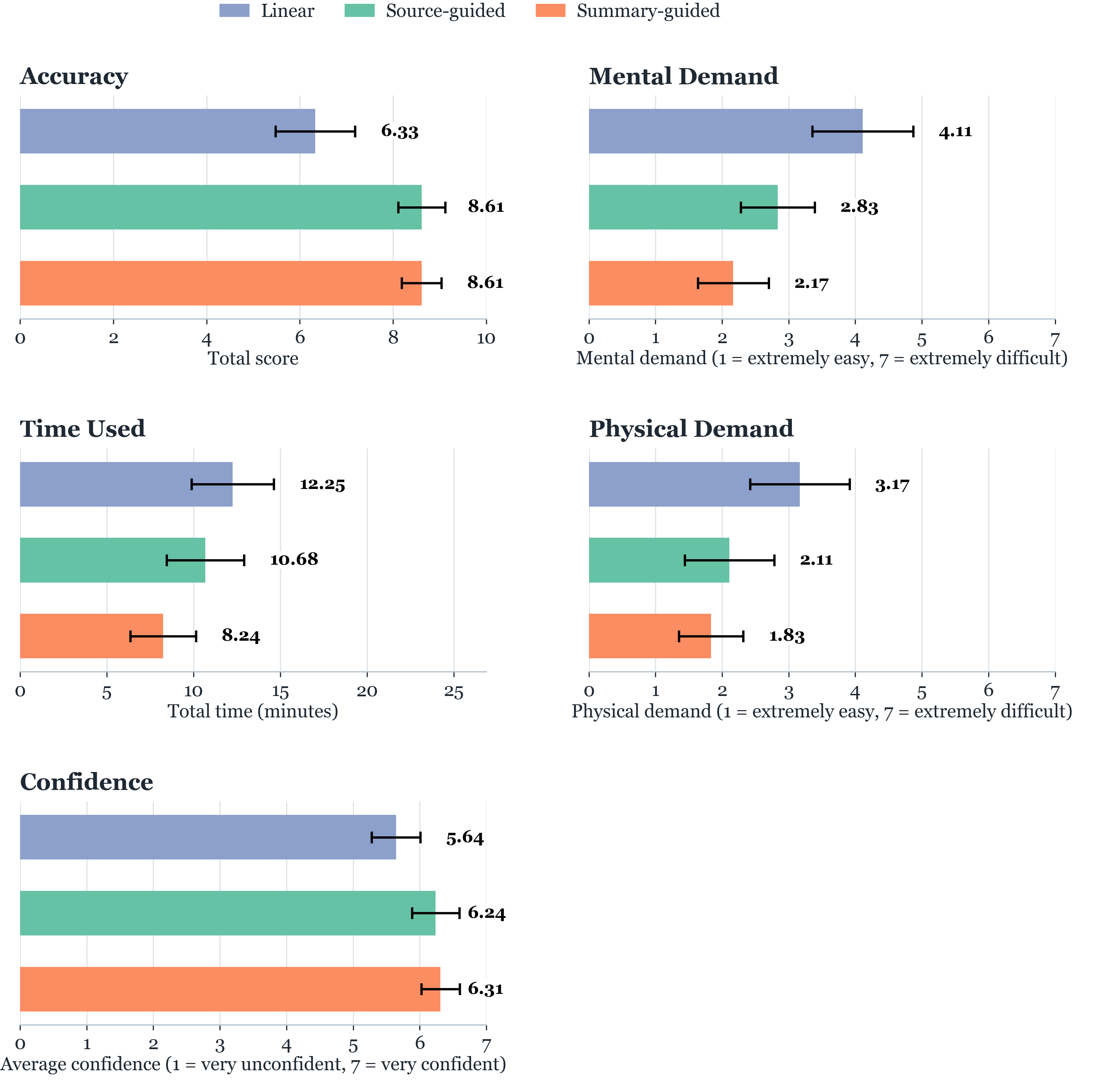}
    \caption{Overall performance across the three document organization conditions, including accuracy, completion time, confidence, mental demand, and physical demand (error bars indicate 95\% confidence intervals). 
    }
    \label{fig:result_01}
    \vspace{-1.5em}
\end{figure}

\subsection{Quantitative Results}
\begin{figure}
    \centering
    \includegraphics[width=\linewidth]{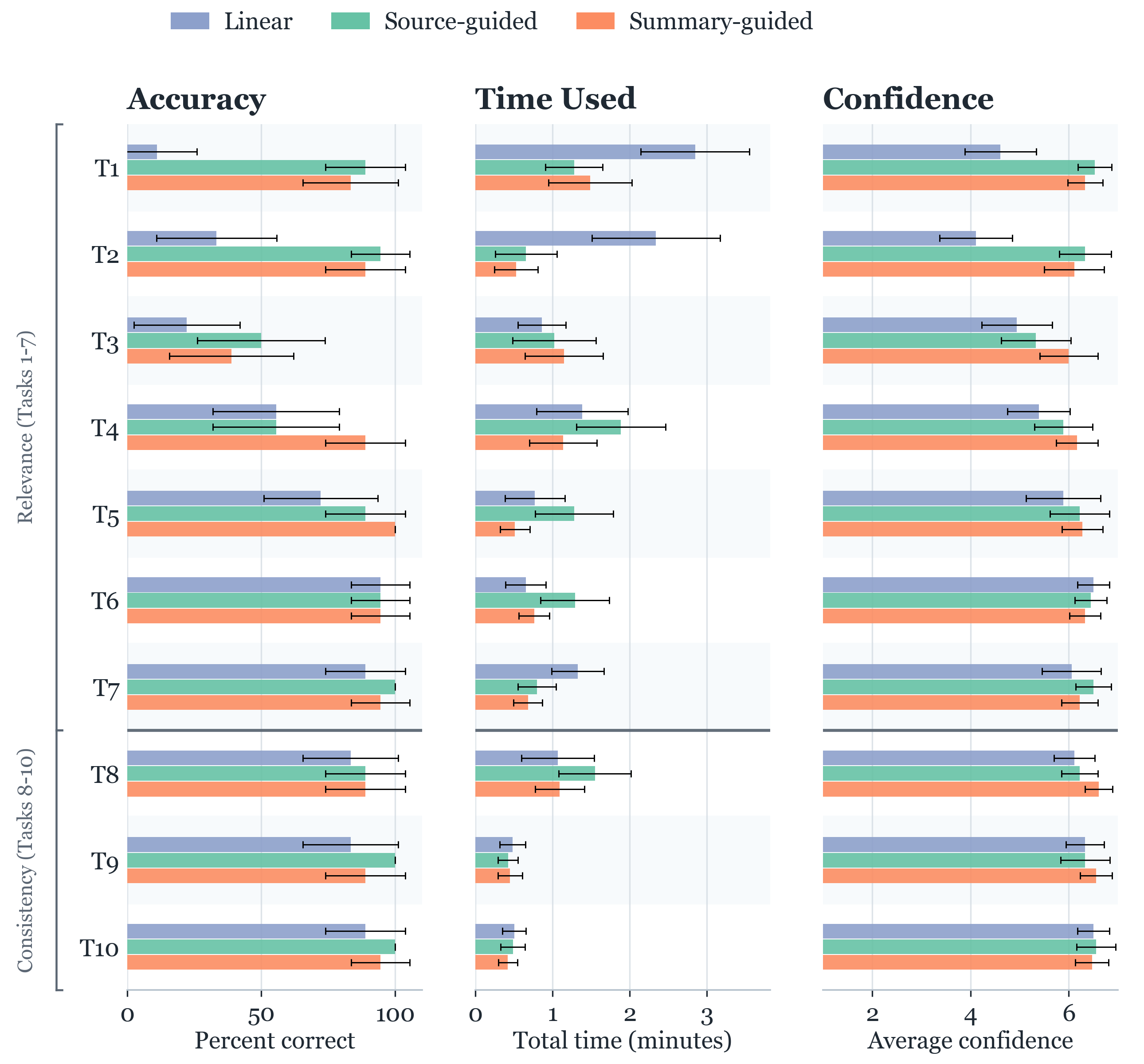}
    \caption{Task-level performance across the ten study tasks (T1–T10) for the three document organization conditions. Accuracy, completion time, and confidence are shown with 95\% confidence intervals.}
    \label{fig:result_02}
    \vspace{-2em}
\end{figure}

We analyzed each metric using linear mixed-effects~\cite{bates2015fitting} models with \textit{condition} as the main fixed effect, \textit{dataset} as a control variable, and participant (\textit{pid}) as a random intercept. For \cref{fig:result_01}, models were fit across all tasks; for \cref{fig:result_02}, the same model was fit separately within each task category. Reported omnibus $p$-values test the overall effect of \textit{condition}, and pairwise $p$-values come from post hoc comparisons of the fitted models.

\paragraph{Overall performance.}
\Cref{fig:result_01} shows the overall effect of interface condition across all tasks. 

For \textit{accuracy}, mean scores were 6.33 for \linear{}, 8.61 for \sourceguided{}, and 8.61 for \summaryguided{}, with a significant condition effect ($p = 8.07\times10^{-8}$). Both guided layouts outperformed \linear{} by 2.28 points (both $p = 2.70\times10^{-10}$), with no difference between them ($p = 1.00$). For \textit{completion time}, means were 12.25, 10.68, and 8.24 minutes, respectively, with a significant condition effect ($p = 2.15\times10^{-4}$). \summaryguided{} was faster than \linear{} by 4.01 minutes ($p = 4.10\times10^{-6}$) and faster than \sourceguided{} by 2.44 minutes ($p = 0.0050$). \sourceguided{} was numerically faster than \linear{}, but not significantly ($p = 0.0716$).

\begin{takeawaybox}
\textbf{Takeaway 1.} Spatial document organization improves the effectiveness of MDS verification, and \summaryguided{} provides the best overall operating point by matching the accuracy gains of both spatial layouts while delivering the largest efficiency benefit.
\end{takeawaybox}

For \textit{confidence}, mean ratings were 5.64, 6.24, and 6.31, respectively, with a significant condition effect ($p = 3.47\times10^{-6}$). Both \sourceguided{} and \summaryguided{} were higher than \linear{} ($p = 9.58\times10^{-7}$ and $p = 3.90\times10^{-8}$), with no difference between the two guided layouts ($p = 0.552$).

Workload ratings showed the same overall pattern. For \textit{mental demand}, means were 4.11 for \linear{}, 2.83 for \sourceguided{}, and 2.17 for \summaryguided{}, with a significant condition effect ($p = 1.05\times10^{-6}$). Both guided layouts were lower than \linear{}, and \summaryguided{} was also lower than \sourceguided{} ($p = 0.030$). For \textit{physical demand}, means were 3.17, 2.11, and 1.83, with a significant condition effect ($p = 8.97\times10^{-4}$). Both guided layouts were lower than \linear{}, while the difference between the two guided layouts was not significant ($p = 0.413$).

\begin{takeawaybox}
\textbf{Takeaway 2.} Guided spatial layouts improve the user experience of MDS verification by making users more confident while reducing perceived workload, with \summaryguided{} providing the lowest mental demand overall.
\end{takeawaybox}

\paragraph{Task-specific results.}~\Cref{fig:result_02} shows Task-level performance across the ten study tasks (T1-T10) for the three conditions.

For \textit{T1} (topic identification), all three metrics showed significant condition effects. Both guided layouts outperformed \linear{} in \textit{accuracy}, with gains of 77.78 percent correct for \sourceguided{} and 72.22 percent correct for \summaryguided{} (both $p < 0.001$). Both guided layouts were also faster than \linear{} in \textit{completion time}, by 1.57 and 1.36 minutes, respectively (both $p < 0.001$), and both produced higher \textit{confidence}, by 1.93 and 1.72 points (both $p < 0.001$). \textit{T2} (topic coverage estimation) showed the same pattern. Relative to \linear{}, \sourceguided{} and \summaryguided{} improved \textit{accuracy} by 61.11 and 55.56 points (both $p < 0.001$), reduced \textit{completion time} by 1.68 and 1.81 minutes (both $p < 0.001$), and increased \textit{confidence} by 2.22 and 2.00 points (both $p < 0.001$).

The remaining relevance tasks showed more selective effects. In \textit{T3} (summary emphasis comparison), only \textit{confidence} had a significant omnibus effect ($p = 0.0064$), with pairwise gains for \sourceguided{} in \textit{accuracy} (+27.78\%, $p = 0.024$) and for \summaryguided{} in \textit{confidence} (+1.06, $p = 0.0017$). In \textit{T4} (most relevant document to the summary), \summaryguided{} improved \textit{accuracy} (+33.33\%, $p = 0.0039$) and \textit{confidence} (+0.78, $p = 0.0072$), while \sourceguided{} was slower (+0.50 minutes, $p = 0.048$). In \textit{T5} (least relevant document to the summary), \summaryguided{} improved \textit{accuracy} (+27.78\%, $p = 0.0048$), whereas \sourceguided{} was slower (+0.51 minutes, $p = 0.020$).

For sentence-level relevance, effects were mostly temporal. In \textit{T6} (most relevant documents to a summary sentence), only \textit{completion time} differed: \sourceguided{} was slower than \linear{} (+0.64 minutes, $p < 0.001$), while \summaryguided{} did not differ ($p = 0.549$). In \textit{T7} (most relevant summary sentence for a document), both guided layouts were faster than \linear{} (-0.53 and -0.65 minutes; $p = 0.0016$ and $p = 1.19\times 10^{-4}$), with no significant effects for \textit{accuracy} or \textit{confidence}.

\begin{takeawaybox}
\textbf{Takeaway 3.} Guided layouts are most beneficial for relevance judgments in MDS verification: both layouts improve topic identification and topic coverage estimation, while \summaryguided{} provides the clearest advantage for judging document relevance to the summary.
\end{takeawaybox}

The consistency tasks showed fewer significant condition effects. For \textit{T8} (supporting documents for a summary sentence), there was no condition effect for \textit{accuracy} ($p = 0.821$) or \textit{completion time} ($p = 0.162$), but there was a significant effect for \textit{confidence} ($p = 0.0135$). Only \summaryguided{} produced higher \textit{confidence} than \linear{}, by 0.50 points ($p = 0.0052$). For \textit{T9} (contradicted claim identification), the omnibus condition effects were not significant for any metric, although \sourceguided{} showed a pairwise \textit{accuracy} gain of 16.67 points over \linear{} ($p = 0.046$). For \textit{T10} (unsupported claim identification), no significant condition effects were observed for \textit{accuracy} ($p = 0.332$), \textit{completion time} ($p = 0.485$), or \textit{confidence} ($p = 0.873$).

\begin{takeawaybox}
\textbf{Takeaway 4.} Consistency tasks are less sensitive to layout condition: unlike the relevance tasks, they show few significant differences across interfaces, with only isolated improvements such as higher confidence for \summaryguided{} in T8 and a modest accuracy gain for \sourceguided{} in T9.
\end{takeawaybox}

\subsection{Qualitative Feedback}
\label{sec:qualitative}

Participant feedback broadly aligned with the quantitative results, with many participants expressing a clear preference for \summaryguided{}. A recurring reason was that the layout made relevance to the summary easy to perceive directly from document position, while still supporting both overview and detailed inspection. Participants repeatedly described this condition as intuitive and easy to navigate.

\begin{compactquote}
``Top-left is most relevant \ldots{} easy to look.''

``The order from top to bottom correspond[s] [to] the summary sentence \ldots{} [it] makes more sense.''

``It has flexibility to let me see the whole and the parts in detail.''
\end{compactquote}

\begin{takeawaybox}
\textbf{Takeaway 5.} Participants preferred \summaryguided{} because document--summary relevance was immediately visible and supported both overview and detailed inspection in a single workflow.
\end{takeawaybox}

Feedback on \sourceguided{} revealed a different strength. Participants who preferred this layout often focused on its ability to reveal relationships within the source collection and provide additional spatial reference points for navigation.

\begin{compactquote}
``Quickly locate relevant document by the layout.''

``It has summary sentence dot, so it give[s] us another reference point to see which documents \ldots{} are close to the summary.''
\end{compactquote}

At the same time, several participants found \sourceguided{} less intuitive for ranking documents by summary relevance.

\begin{compactquote}
``It was not intuitive for me to see which is the most relevant.''

``The document layout was abstract for me to understand.''
\end{compactquote}

\begin{takeawaybox}
\textbf{Takeaway 6.} \sourceguided{} was useful for understanding relationships among source documents, but less clear for judging which documents were most relevant to the summary.
\end{takeawaybox}

In contrast, feedback on \linear{} emphasized the extra effort required to search and compare documents manually. Although some participants appreciated its simple ranked order, comments focused on the additional work needed to navigate the document set.
A few participants also explicitly contrasted the absence of richer interaction support in \linear{} with the guided layouts. 

\begin{compactquote}
``For linear mode(layout), I have to do a lot of searching.''

``The only good thing for the linear one is the linear order of documents that ranked by the relevance.''
\end{compactquote}

\begin{takeawaybox}
\textbf{Takeaway 7.} Compared with \linear{}, the guided layouts reduced manual search and made cross-document relevance relationships easier to understand.
\end{takeawaybox}

\section{Discussion}
\label{sec:discussion}

\paragraph{\textbf{Reliable Use of LLM-generated Summary}}
\oursystem{} supports more reliable use of generated summaries. With \oursystem{}, users can inspect whether a summary covers relevant sources and whether its claims are grounded in the source collection. Guided spatial layouts support relevance verification by showing which documents and topic regions matter for the summary and where source content may be underrepresented, reducing the risk that users miss coverage gaps or misplaced emphasis. Claim-level provenance supports consistency verification by decomposing summary sentences into claims and linking each claim to evidence spans, inference judgment labels, and source context.

Together, these views turn summary verification from open-ended reading into targeted inspection: spatial relevance guidance tells users where to look, while provenance tells them what evidence to judge once they get there. The usage scenarios illustrate how this workflow supports inspection of coverage and evidence, and the user study shows that guided layouts improved verification accuracy, efficiency, confidence, and workload over the \linear{} baseline. Thus, \oursystem{} does not make an LLM summary automatically reliable; instead, it supports more reliable use of generated summaries by making their relevance and grounding easier to verify.

\paragraph{\textbf{Human-in-the-loop Auditing of LLM Outputs}}

\oursystem{} supports human-in-the-loop auditing by placing users at key decision points rather than treating automated outputs as final answers. Users choose what to inspect, expand sentences into claims, compare claims with the original text, and examine each inference judgment label with its rationale, highlighted evidence, and source context. Model outputs serve as cues to question, while users decide whether the summary is supported, contradicted, or under-supported.

\oursystem{} also helps users audit failure cases for LLM-as-a-judge in the verification pipeline. LLM-as-a-judge failures can occur at two stages: claim decomposition and claim-level inference judgment. For claim decomposition, the original summary sentence is shown alongside its generated atomic claims in the summary panel, allowing users to detect decomposition errors by checking whether the claims preserve the sentence's meaning (as shown in \cref{fig:teaser}S1). For the claim-level inference judgment stage, each inference judgment label is linked to the judge's rationale, highlighted evidence spans, and source-document context (as shown in \cref{fig:teaser}C). Users can therefore detect inference-judgment errors by checking whether the cited evidence actually justifies the assigned \textsc{Supports}, \textsc{Contradicts}, or \textsc{Insufficient Evidence} label, and whether the LLM-generated explanation is coherent and aligned with both the label and the visible source evidence. Thus, \oursystem{} does not treat LLM outputs as final ground truth. It presents decomposition and inference-judgment outputs as inspectable and contestable cues while final reliability judgments remain with the user.

\paragraph{\textbf{Comparison with Alternatives}}

\Cref{tab:method_comparison} compares \oursystem{} with representative alternatives for summary verification. Compared with existing visualization systems designed for summary verification such as SummVis~\cite{vig2021summvis} and RTSUM~\cite{cho2024rtsum}, \oursystem{} extends the design space by combining spatial guidance and multi-document support with multi-level provenance visualization.

We also compare \oursystem{} with broader content-generation tools that also provide verification support. Direct LLM prompting can synthesize multiple sources, but lacks spatial guidance and persistent provenance tracking, making relevance and consistency issues hard to detect without repeated prompting and manual cross-checking. Commercial LLM tools such as NotebookLM~\cite{google2026notebooklm} add citations and side-panel source access, but still require users to manually connect summary statements to evidence across documents. We use \linear{} as a controlled baseline for this common commercial provenance pattern, and make it a fair comparison by adding basic highlights for evidence sentences. The user study shows that \oursystem{} outperforms this baseline for relevance and consistency verification, suggesting that spatial organization adds support beyond source access and citations. More details are provided in Appendices~A and~B.

AwesomeLit~\cite{xie2026awesomelit} represents recent research on agent-supported content generation across multiple sources, but provides limited spatial guidance and provenance visualization for verifying generated content. Although it supports efficient and grounded literature review generation, verification of generated content remains less guided than in \oursystem{}.

Across these comparisons, the distinction of \oursystem{} is the integration of spatial guidance, claim-level provenance visualization, and multi-document evidence inspection within one verification workflow.

\begin{table}[t]
\centering
\caption{Comparison of \oursystem{} with representative alternatives.}
\label{tab:method_comparison}
\vspace{-1em}
\scriptsize
\setlength{\tabcolsep}{0.5pt}
\renewcommand{\arraystretch}{0.92}
\begin{tabular}{@{}p{0.34\columnwidth}>{\centering\arraybackslash}p{0.20\columnwidth}>{\centering\arraybackslash}p{0.20\columnwidth}>{\centering\arraybackslash}p{0.20\columnwidth}@{}}
\toprule
\textbf{Method} & \textbf{Spatial guidance} & \textbf{Provenance visualization} & \textbf{Multi-document support} \\
\midrule
SummVis~\cite{vig2021summvis} & \supportlow{} & \supporthigh{} & \supportlow{} \\
RTSUM~\cite{cho2024rtsum} & \supportlow{} & \supporthigh{} & \supportlow{} \\
\midrule
Direct LLM prompting & \supportlow{} & \supportlow{} & \supporthigh{} \\
NotebookLM~\cite{google2026notebooklm} & \supportlow{} & \supportlow{} & \supporthigh{} \\
AwesomeLit~\cite{xie2026awesomelit} & \supportmedium{} & \supportmedium{} & \supporthigh{} \\
\midrule
\linear{} baseline & \supportlow{} & \supportmedium{} & \supporthigh{} \\
\oursystem{} & \supporthigh{} & \supporthigh{} & \supporthigh{} \\
\bottomrule
\end{tabular}
\vspace{-0.25em}
\parbox{0.98\columnwidth}{\scriptsize \textit{Note.} Ratings summarize each method's degree of support for verification workflows: \supportlow{} = low, non-primary support; \supportmedium{} = medium, partial support; \supporthigh{} = high, explicit support.}
\vspace{-2em}
\end{table}

\paragraph{\textbf{Limitation \& Future Work}}
The methods in \oursystem{} have three limitations. First, the verification pipeline partly depends on the underlying LLM, since claim-level inference judgments and evidence links can vary with model quality. However, our goal is not to validate the LLM itself, but to make its verification outputs inspectable so users can check them against source evidence. Second, \oursystem{} focuses on evidence-centered verification through document cards and sentence-level highlighting, and future work could integrate richer document viewers and annotation support. Third, the workflow is primarily user-directed. Future systems could add transparent, contestable agent guidance for summary verification.

\section{Conclusion}

We presented \oursystem{}, a visual analytics system for verifying LLM-generated summaries over multi-document sources. \oursystem{} organizes evidence as a spatial provenance workspace. Its contribution lies in integrating spatial document organization and visual provenance tracking into a coherent workflow for summary verification.

Our study shows that this spatial framing provides measurable benefits over the \linear{} baseline. Both guided layouts increased overall accuracy by 36.0\%, and \summaryguided{} provided the strongest overall performance: it reduced completion time by 32.7\%, increased confidence by 11.9\%, and lowered mental and physical demand by 47.2\% and 42.3\%, respectively. These results suggest that reliable use of generated summaries requires more than access to cited sources, and that verification benefits from visual structures that organize provenance around relevance and consistency checks.

\section*{Supplemental Materials}
\label{sec:supplemental_materials}

Supplemental materials are available through our \href{https://osf.io/6j5c4/overview?view_only=5903d7db4ce24045974f7d8ae911fd26}{\textcolor{blue}{OSF project}}, including the supplemental video and source code. The source code is also available on \href{https://github.com/jiahaoxu14/STT_Attributor}{\textcolor{blue}{GitHub}}, and an interactive demo is available at the \href{https://stt-attributor.vercel.app/}{\textcolor{blue}{project website}}.

\acknowledgments{This material is based upon work supported in whole or in part with funding from the Department of Defense (DoD). Any opinions, findings, conclusions, or recommendations expressed in this material are those of the author(s) and do not necessarily reflect the views of the DoD and/or any agency or entity of the United States Government.}

\bibliographystyle{vendor/ieee-tvcg-2026/abbrv-doi-hyperref}

\bibliography{bibliography/references}

\clearpage
\appendix 
\crefalias{section}{appendix} 
\input{supplemental/appendix-content}

\end{document}

%% file: supplemental/appendix-content.tex
\providecommand{\standaloneappendixheading}{}

\begin{strip}
\centering
    \standaloneappendixheading
    \includegraphics[width=\textwidth]{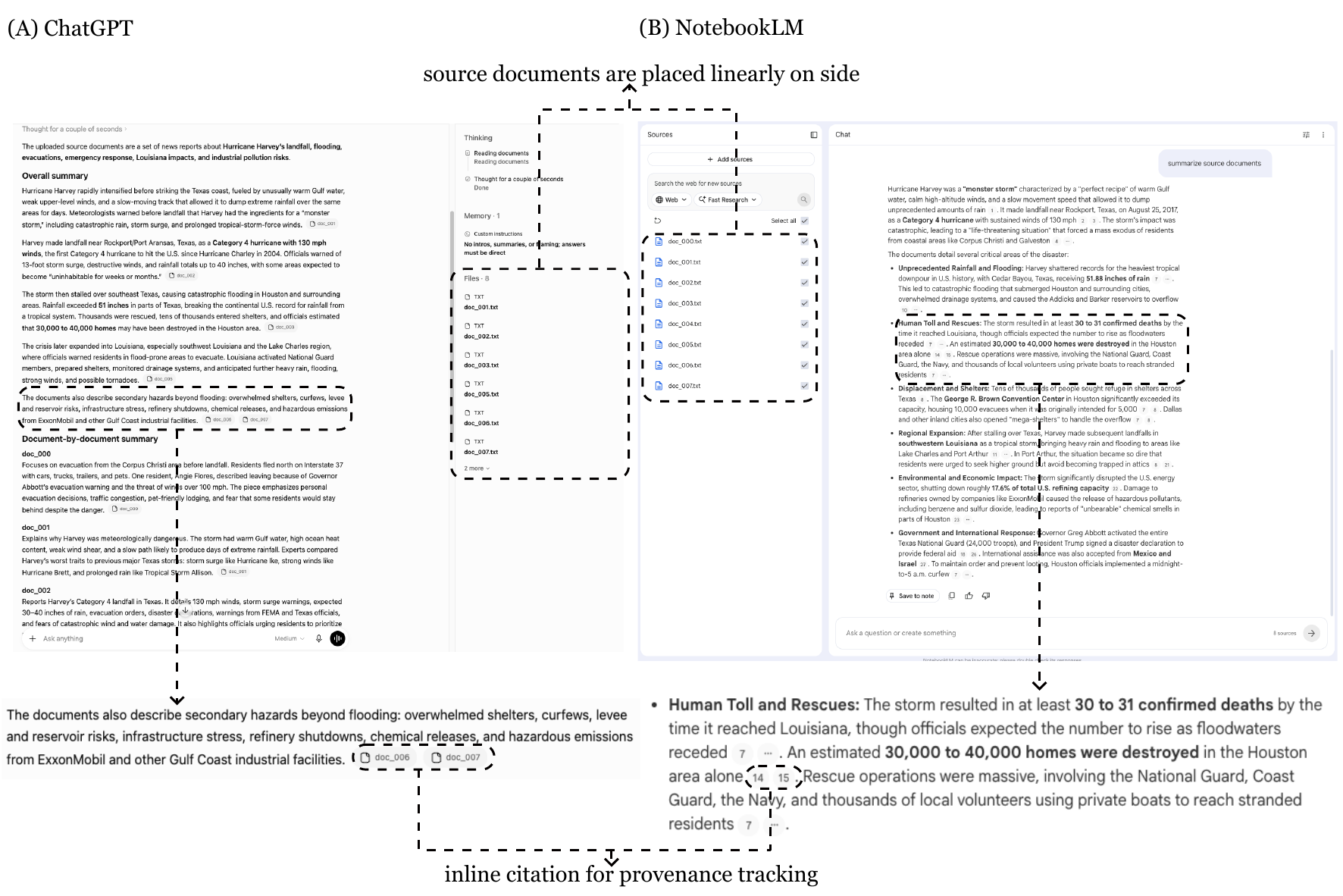}
    \captionof{figure}{Examples of commercial LLM interfaces that expose source materials for verification. (A) ChatGPT and (B) NotebookLM use inline citations in generated text to link summary content to source materials shown linearly in adjacent source views.}
    \label{fig:appendix_commercial_tools}
\end{strip}

\section{Commercial LLM Tools}
\label{app:commercial_tools}

Both examples in \cref{fig:appendix_commercial_tools} illustrate a common provenance pattern in commercial LLM tools. In ChatGPT, generated text appears in the main conversation while uploaded documents are listed in a side panel, and individual statements are linked to source files through inline citation tags. In NotebookLM, the source collection is also shown linearly in a side panel, while numbered citations embedded in the generated answer point users back to source materials. In both cases, provenance is exposed primarily as citation-to-source access.

This design makes source access available during verification, but it does not tell users which parts of the generated summary are most likely to be problematic. A citation may point to a source file, yet the user still has to decide whether the cited context actually supports the summary statement. This is the gap that \oursystem{} targets: it guides verification toward potentially problematic summary spans rather than treating source access as sufficient.

\begin{figure*}[!t]
\centering
    \includegraphics[width=0.9\textwidth]{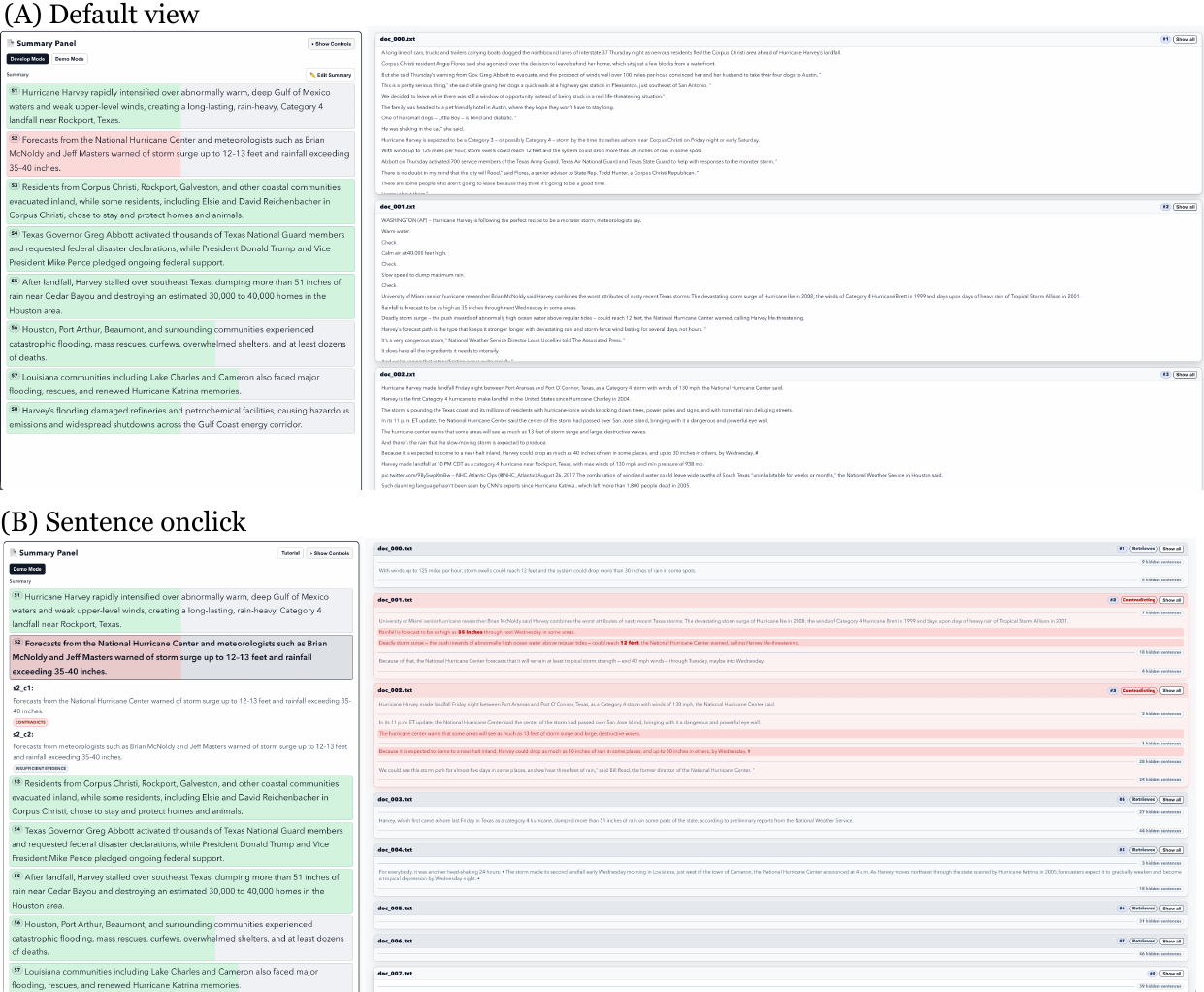}
    \caption{The \linear{} condition used in the user study. The interface follows a familiar linear provenance layout, presenting the generated summary alongside source materials and highlighted evidence sentences.}
    \label{fig:appendix_linear_condition}
\end{figure*}

\section{Linear Condition in User Study}
\label{app:linear_condition}

The \linear{} condition presents source materials as a ranked list, similar to the source lists used in modern commercial LLM tools (\cref{fig:appendix_commercial_tools}). In the default state (\cref{fig:appendix_linear_condition}A), no summary sentence is selected and documents are ordered by aggregate similarity, defined as each document's strongest sentence-level match to any summary sentence. Each row shows the document name, rank, and contents.

After a user clicks a summary sentence (\cref{fig:appendix_linear_condition}B), that sentence becomes the active sentence. The selected summary sentence is visually marked, and its decomposed atomic claims are shown when available. The document list then re-ranks around the selected sentence, moving documents with stronger matching source sentences higher in the list. Document cards also switch to an evidence-focused view: source sentences tied to the selected summary sentence remain visible, support and contradiction evidence is color-highlighted, unrelated context is compressed into hidden-context rows, and each relevant document card scrolls toward its best matching source sentence. Clicking the same summary sentence again clears the selection and returns \linear{} mode to the aggregate default ranking.

We use this condition as the baseline because it captures the dominant provenance pattern in current LLM tools: the generated summary is separated from a linear source list, and verification depends on users following citations back to the source materials. We add basic sentence-level evidence highlights so the baseline provides comparable evidence access, allowing the study to focus on whether spatial guidance improves verification beyond a linear provenance interface.

\section{Prototype Implementation Details}
\label{app:prototype_implementation}

\paragraph{\textbf{System Architecture.}}
The prototype is implemented as a web-based dashboard with a React, TypeScript, and Vite frontend and a Flask backend. The frontend controls the whiteboard workspace and summary panel, and manages the interaction state as users select sentences or claims, inspect evidence, switch layouts, and brush regions. The backend handles document processing and model-facing computation. It uses OpenAI APIs together with NumPy, scikit-learn, UMAP, and spaCy for embedding, projection, evidence retrieval, and related preprocessing.

\paragraph{\textbf{Model Pipeline.}}
For a new document collection, the backend segments source documents into sentences and computes sentence embeddings with \texttt{text-embedding-3-small}. These embeddings support sentence-summary similarity computation, evidence retrieval, and the spatial layouts described in the main paper. The prototype uses \texttt{gpt-5.1} as the default LLM for summary generation, claim decomposition, topic labeling, and claim-level inference judgment. In the judgment step, the model receives retrieved evidence packets and returns a \textsc{Supports}, \textsc{Contradicts}, or \textsc{Insufficient Evidence} label with a rationale that is surfaced in the interface.

\paragraph{\textbf{Runtime and Reuse.}}
No model training or offline preprocessing is required. Users can load a fresh document collection immediately, but they need to wait for the embedding and LLM APIs to respond before the full evidence-aware dashboard is available. Runtime is dominated by API latency and scales with the number of documents, sentences, and atomic claims. The system can also save and reload a complete session, preserving the generated summary, claim structure, inference judgment labels, evidence links, topic corpus, and both \summaryguided{} and \sourceguided{} layouts. When a saved session is loaded, users can begin inspecting the dashboard immediately without rerunning the model pipeline.

\paragraph{\textbf{Code and Model Configuration.}}
Additional implementation details are available in the supplemental source code. The model choices reported above describe the prototype configuration used in our study, but the system architecture is not tied to these specific models. Developers can configure different compatible models for the main LLM and embedding components based on their needs, or expose model selection in the frontend to let users choose models.